\documentclass[11pt]{article}
\topmargin -15mm\usepackage[round]{natbib}
\usepackage{graphicx}
\usepackage{amssymb}
\usepackage{amscd}
\usepackage{mathrsfs}
\usepackage{longtable,lscape}
\usepackage{amsthm}
\usepackage{amsfonts}
\usepackage{amsmath}
\usepackage{bbm}
\usepackage{float}
\usepackage{url}
\usepackage{hyperref}

\newcommand{\captionfonts}{\footnotesize}
\makeatletter 
\long\def\@makecaption#1#2{%
\vskip\abovecaptionskip
\sbox\@tempboxa{{\captionfonts #1: #2}}%
\ifdim \wd\@tempboxa >\hsize
{\captionfonts #1: #2\par}
\else
\hbox to\hsize{\hfil\box\@tempboxa\hfil}%
\fi
\vskip\belowcaptionskip}
\makeatother
\begin{document}
\title{Quantum Bose-Einstein Statistics for 
Indistinguishable \\ Concepts in Human Language}
\author{Lester Beltran\footnote{Center Leo Apostel for Interdisciplinary Studies, Free University of Brussels (VUB), Krijgskundestraat 33,
1160 Brussels, Belgium, lbeltran@vub.ac.be, lestercc21@yahoo.com}}
\date{}
\maketitle
\begin{abstract}
\noindent We investigate the hypothesis that within a combination of a `number concept' plus a `substantive concept',
such as `eleven animals', the identity and indistinguishability present on the level of the concepts,
i.e., all eleven animals are identical and indistinguishable, gives rise to a statistical structure of the Bose-Einstein type similar to how 
Bose-Einstein statistics is present for identical and indistinguishable quantum particles.
We proceed by identifying evidence for this hypothesis by extracting the statistical data 
from the World-Wide-Web utilizing the Google Search tool. By using the Kullback-Leibler divergence method, 
we then compare the obtained distribution with the Maxwell-Boltzmann as well as with the Bose-Einstein distributions 
and show that the Bose-Einstein's provides a better fit as compared to the Maxwell-Boltzmann's. 
\end{abstract}

\section*{Introduction\label{introduction}}
One of the most successful theories in physics is quantum theory. 
The theory explains the behavior of atoms and the particles that constitute them. The same theory is here used to explain cognitive phenomena i.e., concepts, decision making, and language by using models built from it,
within a scientific discipline which has been called quantum cognition and 
quantum information theory \citep{aerts2009,aertsetal2018e,aertsbeltran2020,aertsbroekaertgaborasozzo2013,
aertsgaborasozzo2013,aertsetal2019,haven2018,arguelles2018,busemeyerpothosfrancotrueblood2011,busemeyerbruza2012,dallachiaragiuntininegri2015a,
dallachiaragiuntininegri2015b,havenkhrennikov2013,khrennikov2010,kwampleskacbusemeyer2015,melucci2015,pothosbusemeyer2009}.

The main aim of our investigation is the study of the statistical behavior of concepts of the form of 
a number concept combined with a substantive concept (e.g., {\it Seven Children},{\it Ten People}) in order to show, 
by making use of web retrieval tasks, that the quantum principle of indistinguishability determines this statistical behavior.
Let us explain using a concrete example of why we believe the quantum principle of indistinguishability plays a role in 
such statistical behavior. 

Consider the concept combination {\it Eleven Animals}.
As a concept, each of these eleven animals is perfectly identical and completely indistinguishable from each other one of the eleven animals.
For real physical animals, which bodies occupy each instant of time a part of space, such identity and indistinguishability is never the case.
Even if all the animals are cats, each one of the eleven cats will be different from each other one of the eleven cats.
We hypothesize that if we use the concept combination {\it Eleven Animals} within a piece of text,
the different individuals of the eleven animals will be treated as identical and indistinguishable,
more or less so, depending on the content of the piece of text.

If the text describes in a very concrete way a situation from reality, e.g., a farm where eleven animals are living, 
of which some are cats, some are dogs, and some are horses and cows, then, within that piece of text,
we imagine that we can treat the animals as non-identical and distinguishable, 
because this piece of text describes this real situation.
However, if the piece of text is part of a fictional story, with no real situation in mind,
we imagine the animals to be treated more as identical and indistinguishable.
Like we mentioned already, we will investigate this hypothesis by looking at the statistical behavior of
such combinations of concepts and analyze the nature of the distribution determining this statistical behavior.

In quantum physics, there are two specific distributions, i.e., the Fermi-Dirac's and the Bose-Einstein's. 
The two distributions determine the statistical behavior of, respectively, fermions and bosons, 
fermions being the building blocks of matter, e.g., electrons, protons, 
neutrons and quarks, and bosons being the particles carrying the forces, e.g., photons, W and Z bosons and gluons.

Quantum particles, whether they are fermions or bosons, always behave as identical and indistinguishable entities. 
This means that both the Fermi-Dirac and the Bose-Einstein distributions
describe the statistical behavior of identical particles.
On the contrary, the Maxwell-Boltzmann distribution used for classical particles' behavior
describes a statistical behavior of non-identical and distinguishable entities.

Hence, the method we will follow consists of finding out which type of statistical behavior governs the 
combination of a number concept with a substantive concept,
and more specifically, we will investigate whether one of the quantum distributions 
underlies this behavior or rather the classical Maxwell-Boltzmann distribution.

Investigating corpora of documents like Google Search Engine,
News on Web (NOW) and the Corpus of Contemporary American English (COCA),
we have found evidence of divergence from the Maxwell-Boltzmann distribution and a better fitting of the
Bose-Einstein distribution with the collected data. 

This work is a continuation from what was started in \citet{aertssozzoveloz2015} and \citet{Veloz2015}, 
where already evidence was identified for the Bose-Einstein
statistics providing a better model for the data as compared to the Maxwell-Boltzmann statistics. 
We provide in the present article additional evidence through more extensive and 
reliable data and the explicit use of the Kullback-Leibler divergence method to measure probability distributions' similarity.

In Section \ref{section1}, we briefly introduce the different probability distributions: Maxwell-Boltzmann, Fermi-Dirac and the Bose-Einstein. 
In Section \ref{section2}, we introduce the corpora of documents we investigated to collect data.
In Section \ref{section3}, we explain the method we followed in our analysis, and in Section \ref{section4}, we present our results.
\section{Comparisons of Types of Distributions\label{section1}}
We introduce in this section the probability distributions giving rise to the three types of statistics. 
\subsection{Maxwell Boltzmann Distribution}
The Maxwell Boltzmann Distribution governs classical particles that are distinguishable.
A well-known example of a physical quantity well modeled by the Maxwell-Boltzmann distributions
is the speed of the molecules of an ideal gas. At a given temperature, the molecules have different speeds, 
some are fast, some have moderate speed, some have negligible speed.
The distribution of these speeds is treated as a continuous function because individual speeds are not taken into 
account in a statistical model and anyhow they cannot be measured or predicted.

The Maxwell Boltzmann distribution is used as an approximation to model ideal gas of particles by focusing on 
macroscopic variables such as temperature, pressure, density, or volume.
For particles at thermal equilibrium, it is an excellent approximation.
The distribution was derived by James Clerk Maxwell in 1860, starting from the hypothesis that because of an abundance of molecular collisions, 
such an equilibrium will form\citep{Maxwell(1860a),Maxwell(1860b)}.
In this way, he summarized the properties of a group ofmolecules by applying methods of both probability and statistics.
Ludwig Boltzmann in 1872 \citep{Boltzmann(1872)} also derived the distribution based on 
more explicit but equivalent assumptions and provided a physical basis to it. 
Boltzmann proved that as molecules arrive at equilibrium, they exhibit the Maxwell-Boltzmann distribution.

The Maxwell-Boltzmann distribution also applies for quantum particles whenever the quantum
effects that one particle exerts on another are negligible.
This is why a gas of air molecules at room temperature, although individual molecules are quantum particles, 
can be adequately modeled by the Maxwell-Boltzmann distribution.
Indeed, the mass and velocity of air particles at room temperature are such that the de-Broglie waves of individual air molecules do not overlap,
making the quantum effects of one molecule on another negligible.

The formula of the Maxwell Boltzmann distribution is\\
\begin{equation}
f({E)=}{\frac {1} {{Ae}^{E/kT}}}
\end{equation}\\
where $f(E)$ is the probability density function for a particle to have energy $E$, $A$ is a normalization constant, its value being 
such that all probabilities add up to 1. The exponential in the denominator indicates that the probability for a particle to have a specific energy $E$ gets 
exponentially smaller with increasing energy. The constant $k$ is Boltzmann's constant and $T$ is the absolute temperature. \\
\subsection{Fermi-Dirac Distribution}
The Fermi-Dirac distribution is the quantum distribution that describes the behavior of fermions,
which are quantum particles, hence they obey the quantum principle of indistinguishability. 
The general spin statistics theorem of quantum theory proves that fermions are quantum particles with half integer-spin\citep{Saunders(2009)}.
They are the particles that constitute matter, such as electrons, protons, neutrons, and quarks.
They also obey the Pauli Exclusion Principle, stating that identical fermions cannot be in the same quantum state.
This principle was formulated by the Austrian physicist Wolfgang Pauli in 1925\citep{Saunders(2009)} at first only for electrons.
In 1939, Markus Fierz put forward a formulation of the spin-statistics relation\citep{Fierz(1939)}
making it possible for Wolfgang Pauli to derive the general spin-statistics theorem in 1940
and extended the Pauli Exclusion Principle to all fermions.\citep{Pauli(1940)}.

The Fermi-Dirac distribution was named after Enrico Fermi and Paul Dirac, who worked
on this statistical issue specifically for the electrons and was later extended to all fermions.
For electrons, the Pauli Exclusion Principle prevents two of them with the same spin to occupy the same energy level.
Note that the above mentioned quantum properties of fermions are the origin of matter
having both stability and volume\citep{DysonLenard(1967),LenardDyson(1968),Lieb(1976),MuthapornManoukian(2004)}.

The formula of the Fermi-Dirac Distribution is given by
\begin{equation}
f(E)=\frac{1}{e^{(E-E_{f})/kT}+1}
\end{equation}\\
where $f(E)$ is the probability density function for a particle to have energy $E$, $A$ is a normalization constant, its value being 
such that all probabilities add up to 1, $k$ is the Boltzmann constant, $T$ the absolute temperature, $E_f$ is called the chemical potential and 
$+1$ is the quantum difference because of indestinguishability of particles.
\subsection{Bose-Einstein Distribution}
The Bose-Einstein Distribution governs bosons which are identical and indistinguishable quantum particles with integer spin\citep{Borrelli(2009)}. 
Contrary to fermions, bosons do not obey the Pauli exclusion principle\citep{Li(2006)}. 
Like we mentioned already, if quantum effects are negligible,
the Maxwell-Boltzmann distribution is a good approximation,
which can be shown to be the case for boson gases at high temperatures and low concentrations.

The coming into existence of Bose-Einstein Distribution and it's use for the modeling of bosons,
is interesting from a historical perspective were both Satyendra Nate Bose and
Albert Einstein played an important role \citep{Wali(2006)}.
The Bose-Einstein Distribution as it appears specifically in the Planck radiation law
was a rather accidental discovery of Sathyendra Nate Bose while he was teaches at The University of Dhaka.
The subject of the lecture was the ultraviolet catastrophe and the radiation theory of light, 
While Bose aimed to demonstrate the inadequacy of the pre-Planck theory and how 
the predicted results do not agree with the results of the experiments he committed a simple calculation mistake, 
The mistake turned out to be equivalent to applying a new type of statistics to the quanta of light which later would be called the Bose-Einstein statistics. 
Bose was amazed to see that this mistake made it possible to directly derive Planck's radiation law and hence 
the idea was born that possibly the Maxwell-Boltzmann statistics was not the one to apply to the quanta of light. 
Bose send his findings written down in an article directly to Albert Einstein who was very enthusiastic about the new idea, 
and translated Bose's article in German, and the article was published with an end comment by Einstein \citep{bose1924}. 
The formula of the Bose-Einstein Distribution is given by\\
\begin{equation}
f({E)=}{\frac {1} {{Ae}^{E/kT}-1}}
\end{equation}\\
where $f(E)$ is the probability density function for a particle to have energy $E$, $A$ is a normalization constant, its value being 
such that all probabilities add up to 1, $k$ is Boltzmann's constant, $T$ is the universal temperature 
and $-1$ is the quantum difference because of indistinguishability of particles.

\section{The Documents and Corpus \label{section2}}
A Web Search Engine or Internet Search Engine is a software system or a program specifically
used for searching the web using words stated in the search box.
The search results are presented systematically in an arranged list of web links that are related to 
the specific topic specified by the words used in the search.

Several documents and a search engine were used for our investigation to obtain the frequencies in the combination of words.
The primary search engine used is the well known Google Search Engine,
which has an extensive index of hundreds of terabytes of information from webpages.
Our investigation uses Google.com, the world-wide version, and not the version specific to a country.
Only Google search engines provide the verbatim search (literal word) and show the total frequency of results.
If the literal search is not used, Google will use it's Artificial Intelligence to search instead for associated words.
Similar search engines such as Yahoo and BING! (both of which are owned by the same company), do not give verbatim search.
Other search engines also do not give rise to verbatim search, as the user's profile is also taken into account in the search by the AI. 

Concerning the corpora of text, we used Google Books, which is available for free at https://googlebooks.byu.edu/x.asp.
We also used News on Web (NOW), freely available at https://corpus.byu.edu/now
and the Corpus of Contemporary American English (COCA), freely available at https://corpus.byu.edu/coca.
Google Books is the most extensive accessible corpus, with 155 billion words of books scanned by Google.
Then comes the NOW corpus with 11.1 billion words of texts from news and
periodicals, and finally, COCA has 1 billion words of texts of the story type.
\section{The Method \label{section3}}
\begin{figure}[H]
\begin{center}
\includegraphics[width=0.75\textwidth]{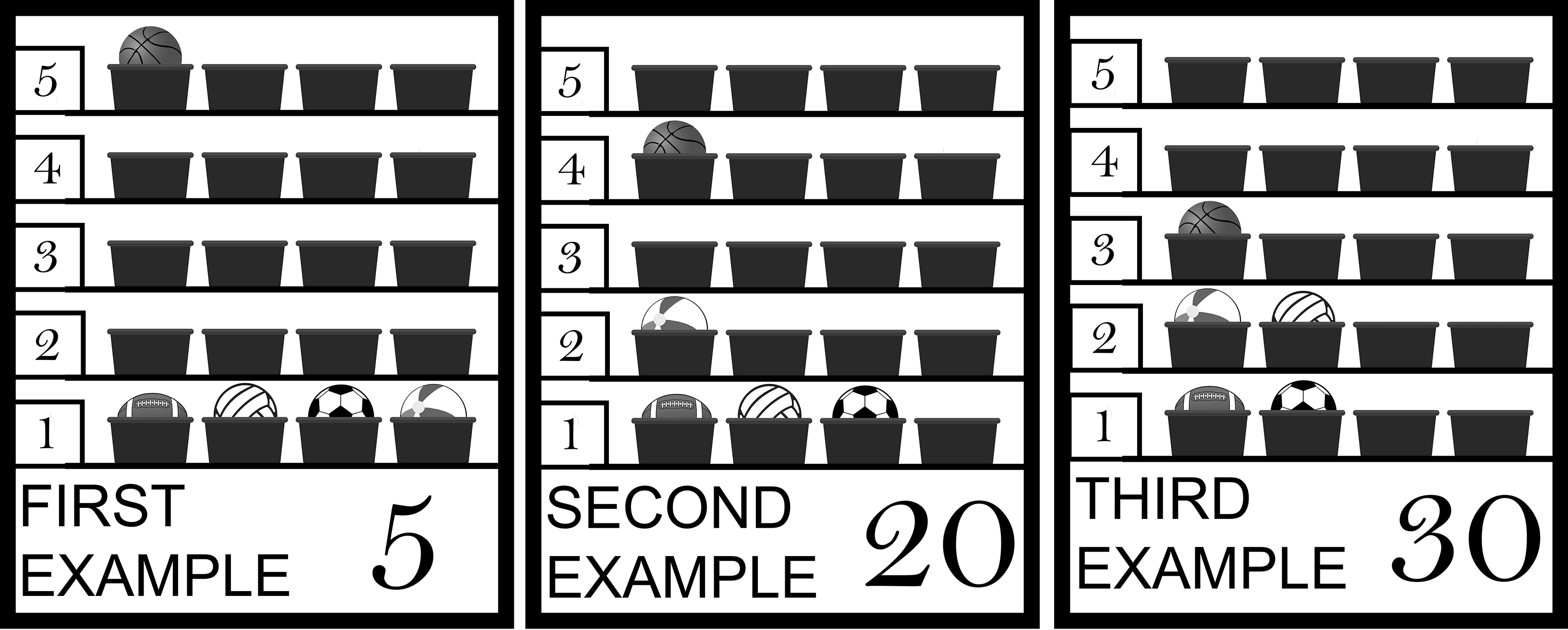}
\caption{An example of the Maxwell-Boltzmann statistics of balls of different types distributed over baskets arranged in different levels}
\label{fig:MBSample}
\end{center}
\end{figure}

We illustrate the Maxwell-Boltzmann statistics making use of 
the example of five balls of different types to be distributed over multiple baskets, 
which are placed four by four on five distinct levels (see  Fig.\ref{fig:MBSample}).
\noindent
The balls are of five different types and hence distinguishable from each other.
Consider as a first example is the situation where one ball is placed on the highest level, and four balls on the lowermost level.
The single ball on the top, by the reason of being distinguishable from each other,
can be replaced by any other distinguishable balls at the bottom.
All of the balls can be replaced with other types of balls, 
provided they follow the same form, i.e., one ball at the topmost and four balls at the bottom.
It is easy to check that the situation shown in the first example can be swapped positions with each other five 
times (see  Fig.ref{Maxwell-Boltzmann Example}). 
For the situation shown in the second example of Fig.\ref{fig:MBSample}, we can re-arrange the balls twenty times.
For the situation shown in the third example of Fig.\ref{fig:MBSample},  we can re-arrange the balls without changing the pattern thirty times. 

The general formula, giving the number of possible re-arrangements, is 
\begin{equation}
{N! \over n_1! \cdot n_2! \cdot n_3! \cdot n_4! \cdot n_5!}
\end{equation}
where $N$ is the total number of balls and $n_i$ is the number of balls at level $i$. We indeed have for the first example
\begin{equation}
{5! \over 4! \cdot 0! \cdot 0! \cdot 0! \cdot 1!} = {1 \cdot 2 \cdot 3 \cdot 4 \cdot 5 \over 1 \cdot 2 \cdot 3 \cdot 4} = 5
\end{equation}
for the second example
\begin{equation}
{5! \over 3! \cdot 1! \cdot 0! \cdot 1! \cdot 0!} = {1 \cdot 2 \cdot 3 \cdot 4 \cdot 5 \over 1 \cdot 2 \cdot 3 } = 4 \cdot 5 = 20
\end{equation}
and for the third example
\begin{equation}
{5! \over 2! \cdot 2! \cdot 1! \cdot 0! \cdot 0!} 
= {1 \cdot 2 \cdot 3 \cdot 4 \cdot 5 \over 1 \cdot 2 \cdot 1 \cdot 2} 
= {3 \cdot 4 \cdot 5 \over 1 \cdot 2} = 30
\end{equation}

\begin{figure}[H]
\begin{center}
\includegraphics[width=0.75\textwidth]{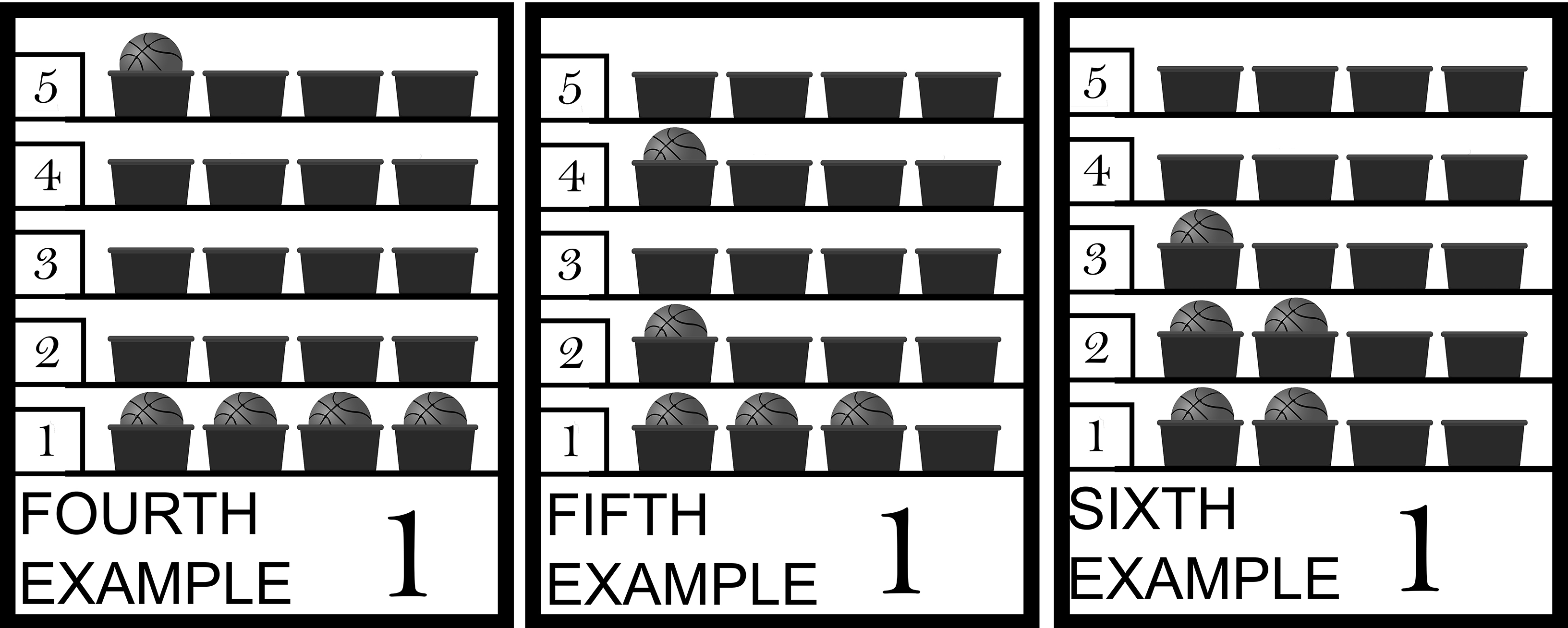}
\caption{An example of the Bose-Einstein statistics of balls of identical types distributed over baskets arranged in different levels}
\label{fig:BESample}
\end{center}
\end{figure}
\noindent
We will now illustrate the Bose-Einstein statistics making use of the example of five balls of the same types to be distributed over multiple baskets, 
which are also placed four by four on five distinct levels (see  Fig.\ref{fig:BESample}).
The balls are all of the same types and hence indistinguishable from each other.
Consider as a first example is the situation where one ball is placed on the highest level, and four balls on the lowermost level.
If the single ball on the top, as a consequence of all balls being indistinguishable from each other,
is replaced by any other indistinguishable balls at the bottom, this will not be apparent.
It is indeed impossible to determine if the balls are swapped and how many times.
For this reason, all arrangements will be identical, which means that there is only a single configuration possible for all three of the previous examples. 
This is the typical situation of the Bose-Einstein statistics.
Also for the situation shown in the second example of Fig.\ref{fig:MBSample} and the third example of Fig.\ref{fig:MBSample} only one configuration is possible.
Since, we already explained the two distributions by examples, 
We will look to different situations of combination of concepts that fit in the above archetypical one of baskets and balls, 
intending to find out whether it is the Maxwell-Boltzmann distribution or the Bose-Einstein distribution, which best models the data.
For example, consider the combination 
`eleven animals' as being the equivalent of our baskets and the two states where an animal can be a cat or a dog 
(in the balls' example `states' have been called `levels'). 

We can then consider the following situations 12 configurations: 
`eleven cats', `ten cats and one dog', `nine cats and two dogs', \ldots, `one cat and ten dogs', `eleven dogs'.  
The Maxwell-Boltzmann distribution for this situation is then 
given by 12 configurations, with respectively the following numbers of distinguishable cases in each configuration: 
\begin{equation}
\begin{gathered}
{11! \over 11! \cdot 0!}= {1}\;\;\;\;\;
{11! \over 10! \cdot 1!} = {11}\;\;\;\;\;
{11! \over 9! \cdot 2!} = {55}\;\;\;\;\;
{11! \over 8! \cdot 3!} = {165}\;\;\;\;\;
{11! \over 7! \cdot 4!} = {330}\;\;\;\;\;
{11! \over 6! \cdot 5!} = {46}\\\\
{11! \over 5! \cdot 6!} = {462}\;\;\;\;\;
{11! \over 4! \cdot 7!} = {330}\;\;\;\;\;
{11! \over 3! \cdot 8!} = {165}\;\;\;\;\;
{11! \over 2! \cdot 9!} = {55}\;\;\;\;\;
{11! \over 1! \cdot 10!} = {11}\;\;\;\;\;
{11! \over 0! \cdot 11!} = {1}
\end{gathered}
\end{equation}
\\\\
We collected the frequency of appearances to check for the presence of MB or BE statistics.
\section{Word Combination \label{section4}}
We considered four corpora in this investigation, Google Search Engine, Google Books,
News on Web (NOW), and Corpus of Contemporary American English (COCA), but three of them, Google Books, NOW and COCA, turned out to be too small 
that most of the searched texts yielded no results.
Only the Google Search Engine gave rise to enough results.
These results are trustworthy since the frequencies are in the range of one to 150 items, 
hence it was possible to verify the results page by page, to check that each of them truthfully contained the specified search text.
For example, in Table \ref{tab:fifteen children}, we consider the concepts {\it Fifteen} and {\it Children}, in the combination {\it Fifteen Children}.
We then consider two possible states for each child, boy and girl. This means that we consider the following 16 strings: 
`fifteen boys and zero girls,' `fourteen boys and one girl,'`thirteen boys two girls', {\ldots}, 
`eight boys and seven girls', `seven boys and eight girls', {\ldots}, `one boy and fourteen girls', `zero boys and fifteen girls'.

Let us first derive the Maxwell-Boltzmann and the Bose-Einstein distributions for this type of situation, 
where we have $N+1$ configurations (strings) of $N$ concepts (Children) and two states (boy and girl), 
of which $n_1$ in the first state and $n_2$  in the second state, such that $N = n_1 + n_2$. 
For Maxell-Boltzmann, the number of distinguishable cases is given by
\begin{equation}
{N! \over n_1! \cdot n_2!}
\end{equation}
We have that
\begin{equation}
\sum_{n_1=0}^N {N! \over n_1! \cdot n_2!} = 2^N
\end{equation}
which means that we have a normalization factor of $1 / 2^N$ and the Maxwell-Boltzmann distribution for this situation is given by
\begin{equation}
f_{MB}(n_1) = {N! \over n_1! \cdot n_2!} \over  2^N
\end{equation}
In the case of Bose-Einstein statistics each one of the $N+1$ strings is equally probable, 
which means that the Bose-Einstein distribution is given by
\begin{equation}
f_{BE}(n_1) = {1 \over N+1}
\end{equation}
However we have to adapt these distributions to take into account the following. 
A search of the string `fifteen boys and zero girls' and the string `zero boys and fifteen girls' cannot be treated in a 
Google search at the same level as the search of the string `one boy and fourteen girls,' or `eight boys and seven girls', 
because `zero girls' and `zero boys' are just not expressed in language.

To remedy this inconvenience, we decided that for each test, we drop
the first (fifteen boys and zero girls) and the last data point (zero boys and fifteen girls).
We can do this without inconveniences since 
we still have fourteen data points left that accurately operate at the same level for a Google search. 
Of these fourteen data points, we will analyze whether they obey the Maxwell-Boltzmann statistics or the Bose-Einstein statistics.

With the above explained correction consisting of dropping the first and last data points for the Maxwell-Boltzmann and Bose-Einstein distributions, 
we have to compare the following data. 
For $1 \le n_1 \le N-1$ we have
\begin{equation}
f_{MB}(n_1) = {{N! \over n_1! \cdot n_2!} \over \sum_{n_1=1}^{N-1}{N! \over n_1! \cdot n_2!}} 
\end{equation}
for the Maxwell-Boltzmann distribution, and for $1 \le n_1 \le N-1$ we have
\begin{equation}
f_{BE}(n_1) = {1 \over N-1}
\end{equation}
for the Bose-Einstein distribution.

The obtained number of appearance (Text Freq. in the table) are presented in Table \ref{tab:fifteen children}.
\begin{table}[h!]
\begin{tabular}{|l|l|l|l|l|l|l|}
\hline
Nos. &Text Combination&Text Freq.&Text Prob.&MB Freq.& MB Prob.&BE Prob.\\ \hline
1.& fifteen boys & 161 & & 1 & & \\ \hline
2.& fourteen boys and one girl & 55 & 0,043205027 & 15 & 0,000457792 & 0,071428571 \\ \hline
3.& thirteen boys and two girls & 49 & 0,038491752 & 105 & 0,003204541 & 0,071428571 \\ \hline
4.& twelve boys and three girls & 62 & 0,048703849 & 455 & 0,013886346 & 0,071428571 \\ \hline
5.& eleven boys and four girls & 99 & 0,077769049 & 1365 & 0,041659037 & 0,071428571 \\ \hline
6.& ten boys and five girls & 155 & 0,121759623 & 3003 & 0,091649881 & 0,071428571 \\ \hline
7.& nine boys and six girls & 154 & 0,120974077 & 5005 & 0,152749802 & 0,071428571 \\ \hline
8.& eight boys and seven girls & 180 & 0,141398272 & 6435 & 0,196392602 & 0,071428571 \\ \hline
9.& seven boys and eight girls & 157 & 0,123330715 & 6435 & 0,196392602 & 0,071428571 \\ \hline
10.& six boys and nine girls & 150 & 0,117831893 & 5005 & 0,152749802 & 0,071428571 \\ \hline
11.& five boys and ten girls & 99 & 0,077769049 & 3003 & 0,091649881 & 0,071428571 \\ \hline
12.& four boys and eleven girls & 70 & 0,054988217 & 1365 & 0,041659037 & 0,071428571 \\ \hline
13.& three boys and twelve girls & 13 & 0,010212097 & 455 & 0,013886346 & 0,071428571 \\ \hline
14.& two boys and thirteen girls & 29 & 0,022780833 & 105 & 0,003204541 & 0,071428571 \\ \hline
15.& one boy and fourteen girls & 1 & 0,000785546 & 15 & 0,000457792 & 0,071428571 \\ \hline
16.& fifteen girls & 169 & & 1 & & \\ \hline
&TOTAL & 1273 & & 32766 & & \\ \hline
\end{tabular}
\caption{The three probability distributions for the example of fifteen children}
\label{tab:fifteen children}
\end{table}
\noindent
We did also retrieve the frequency of the first and last datapoints, `fifteen boys' and `fifteen girls'. 
The first gives 161 results, so that it can be seen that it is of another order of magnitude than the second data point, 
`fourteen boys and one girl', which gives 55, but as we said, it is not used in our modeling. 

The sum of all the frequencies of the fourteen considered datapoints search strings is equal to 1273. 
Let us calculate now their corresponding probabilities so that we can deduce the distribution function.
To compute each search string's probability, 
the frequency of the search string is divided by the sum of  the frequencies of all searched strings except the first and last ones. 

To compute the Maxwell-Boltzmann frequency, $n$ is the total number of search
strings (which is 15) minus one. The $r$ is the $nth$ position of the search text in the
list minus one, e.g., if it is the first search string, it is then $r=(1-1)$, the second search
string is $r=(2-1)$, and so forth. The formula used is

\begin{equation}
C = \frac{n!}{r!*(n-r)!}
\end{equation}\\
Finally, to obtain the Bose-Einstein frequency of a searched string, 
we divide one by the total number of strings (which is 16) minus 2 (because we do not include the first and the last). 
This gives the constant value (1/[16-2]) for all searched strings .

Now that we obtained all of the probabilities, we will need to determine which of the two distributions best models the data.
We have made use of the Kullback-Leibler divergence to identify the best fit.
The Kullback-Leibler divergence is a measurement of how one probability distribution differs from a second probability distribution. 
It is a comparison of the probability distribution of the first event to the probability distribution of the second event \citep{Brownlee(2020)}.
First we compare the search string with the Maxwell-Boltzmann and get their
Kullback-Leibler divergence value.
Let us call the probability distribution of the search string 'S' and the Maxwell-Boltzmann probability distribution 'M' . To obtain the search string (SS) versus Maxwell-Boltzmann (MB) Kullback-Leibler divergence value, we use this formula
\begin{equation}	
KL(SS\|MB) = S*Log(S/M)
\end{equation}
After obtaining the individual SS-MB Kullback-Leibler divergence value, we add them to obtain the total SS-MB Kullback\-Leibler divergence value.
We then compare the search string's probability distribution with the Bose-Einstein probability distribution and get their Kullback-Leibler divergence value.
Let us call the probability distribution of Bose-Einstein 'B'.
To obtain the SS-BE Kullback-Leibler divergence value, we use this formula
\begin{equation}			
KL(SS\|BE) = S*Log(S/B)
\end{equation}
After we obtained the individual SS-BE Kullback\-Leibler divergence value, we add them to get the total SS-BE Kullback\-Leibler divergence value.
To compare the Kullback-Leibler divergences of the data distribution with the Maxwell-Boltzmann and the data distribution with the Bose-Einstein,
we divide the total of SS-MB Kullback\-Leibler divergence value by the total SS-BE Kullback\-Leib divergence value.
The more the number that results is bigger than 1,
the more the Bose-Einstein distribution fits better with the data distribution than the Maxwell-Boltzmann distribution.
The results of some of the examples can be found below.

\begin{figure}[H]
\centering
\includegraphics[width=0.75\textwidth]{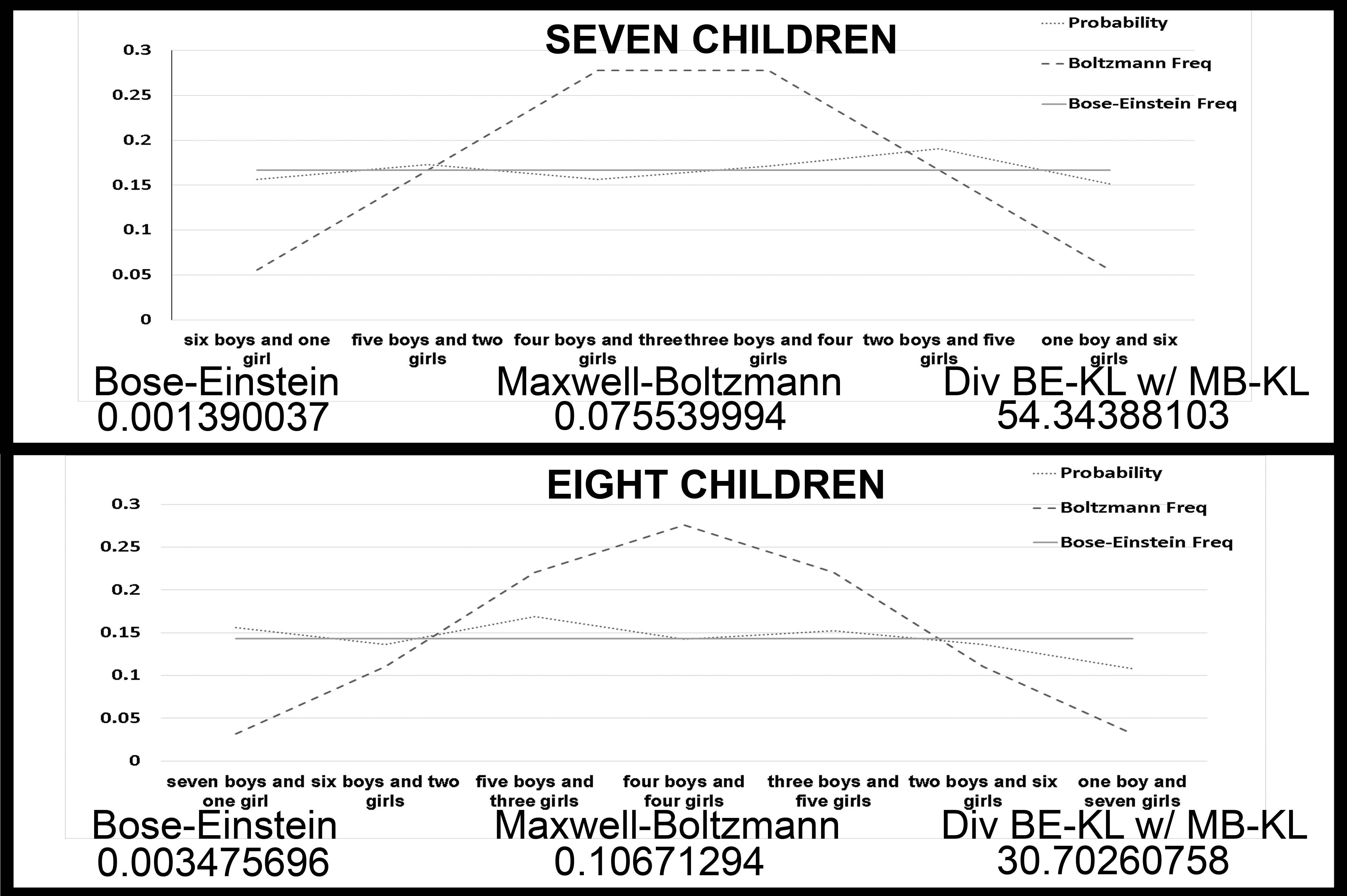}
\caption{Combination of childrens; boys and girls, showing both the SS-MB Kullback-Leibler divergence value and SS-BE Kullback-Leibler divergence value}
\end{figure}
\begin{figure}[H]
\centering
\includegraphics[width=0.75\textwidth]{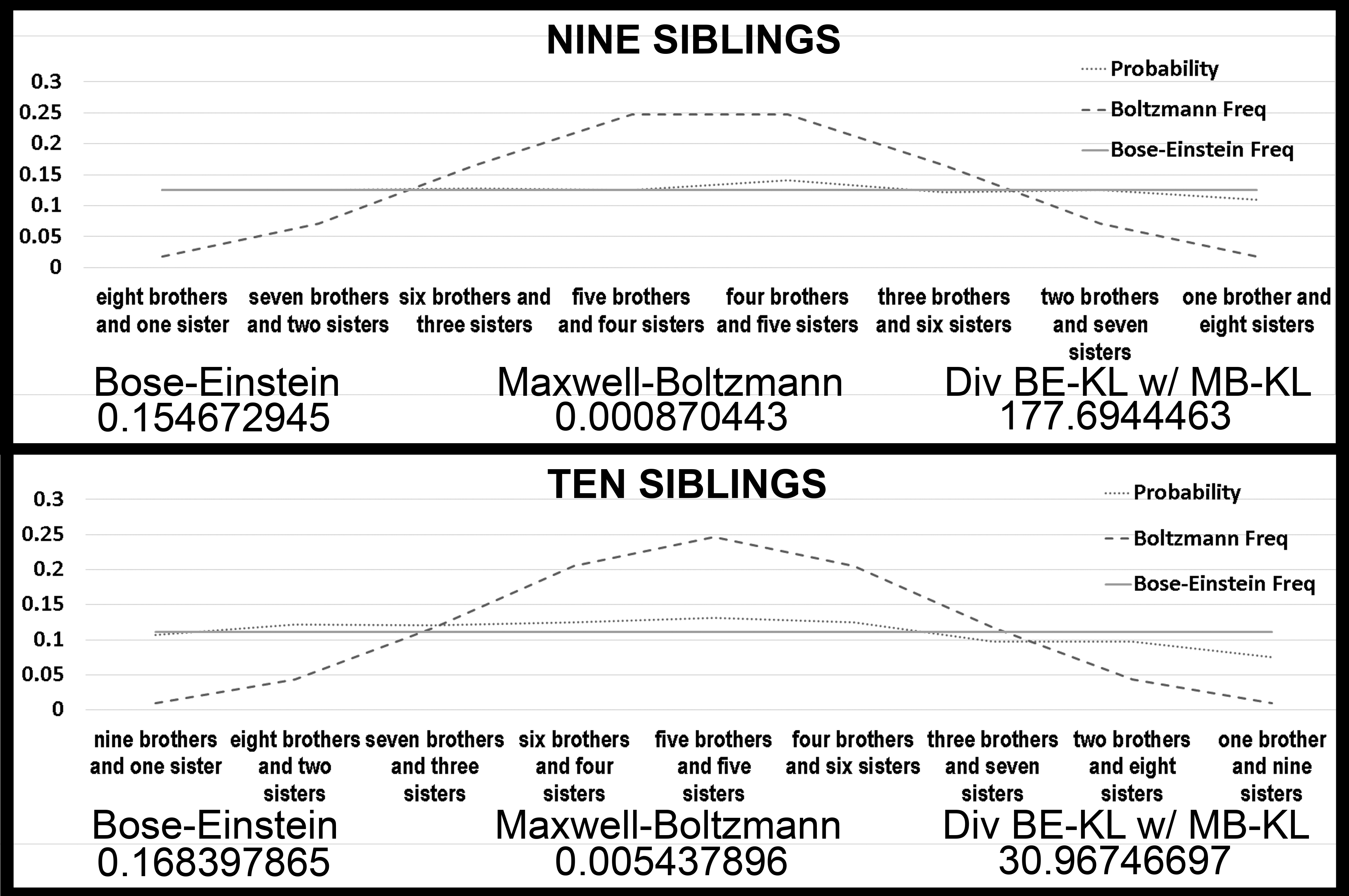}
\caption{Combination of siblings; brothers and sisters, showing both the SS-MB Kullback-Leibler divergence value and SS-BE Kullback-Leibler divergence value}
\end{figure}
\begin{figure}[H]
\centering
\includegraphics[width=0.75\textwidth]{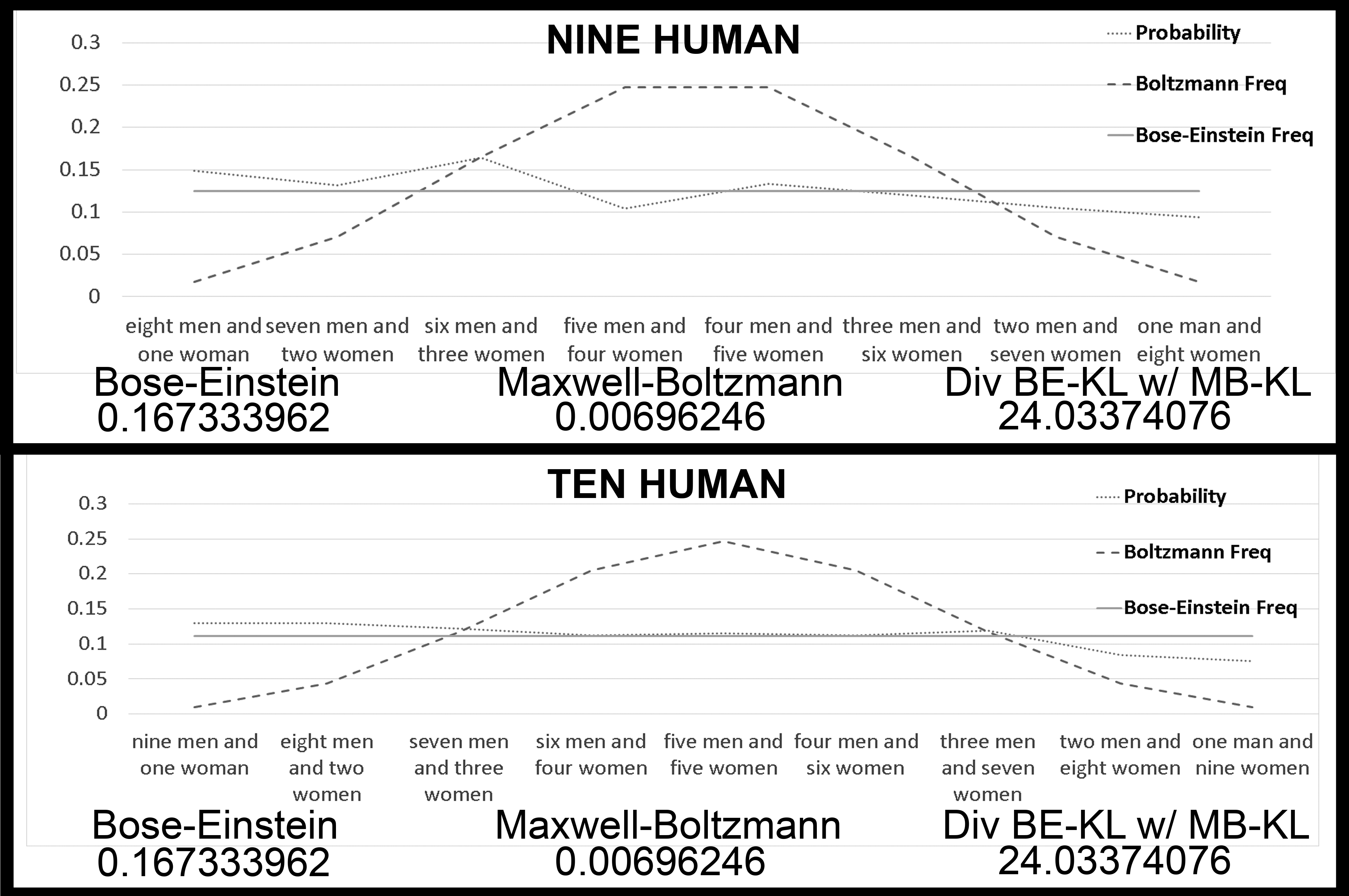}
\caption{Combination of humans; men and women, showing both the SS-MB Kullback-Leibler divergence value and SS-BE Kullback-Leibler divergence value}
\end{figure}
\noindent
The used corpus is large enough to provide reliable frequencies of results.
The results collected show a strong preference for the BE statistics.
The Kullback-Leibler divergence gives a higher value for the Maxwell-Boltzmann distribution
as compared to the value is gives for the Bose-Einstein distribution.
This means that the Bose-Einstein distribution offers a better model for the data than the Maxwell-Boltzmann distribution.The results presented in this article were obtained before we engaged in a more direct investigation of the statistical structure of human language with respect to the basic hypothesis of the way concepts behave similarly to identical quantum particles.
This more direct investigation of the statistical structure of pieces of text that represent stories, short stories as well as stories the size of a novel,
confirms in a very strong way that it is the Bose-Einstein distribution that is at the basis of this
statistical structure and not the Maxwell-Boltzmann distribution \citep{aertsbeltran2020}.

\bigskip
\noindent {\bf Acknowledgements}

\smallskip
\noindent
I thank Massimiliano Sassoli de Bianchi, Tomas Veloz and Diederik Aerts to read over the text of this article and give me their comments and suggestions. 
This work is funded by the European Union's Horizon 2020
Research and innovation programme under the
Marie Sklodowska-Curie grant agreement No:721321.

\end{document}